\documentclass{article}

\usepackage{microtype}
\usepackage{graphicx}
\usepackage{booktabs} 
\usepackage{soul}
\usepackage{url}

\usepackage{hyperref}

\usepackage[accepted]{icml2024}

\usepackage{amsmath}
\usepackage{amssymb}
\usepackage{mathtools}
\usepackage{amsthm}

\usepackage{graphicx,wrapfig,lipsum,booktabs}
\usepackage{booktabs}       
\graphicspath{ {./images/} }
\usepackage{amsmath,bm}
\usepackage{caption}
\usepackage{subcaption}
\usepackage{dsfont}
\usepackage[export]{adjustbox}
\usepackage{multirow}
\usepackage{dblfloatfix}    

\usepackage[capitalize,noabbrev]{cleveref}

\theoremstyle{plain}

\theoremstyle{definition}

\theoremstyle{remark}

\usepackage[textsize=tiny]{todonotes}

\icmltitlerunning{Improving Antibody Humanness Prediction using Patent Data}

\begin{document}

\twocolumn[
\icmltitle{Improving Antibody Humanness Prediction using Patent Data}



\icmlsetsymbol{equal}{*}

\begin{icmlauthorlist}
\icmlauthor{Talip Uçar}{yyy}
\icmlauthor{Aubin Ramon}{comp}
\icmlauthor{Dino Oglic}{yyy}
\icmlauthor{Rebecca Croasdale-Wood}{sch}
\icmlauthor{Tom Diethe}{yyy}
\icmlauthor{Pietro Sormanni}{comp}
\end{icmlauthorlist}

\icmlaffiliation{yyy}{Centre for AI, BioPharmaceuticals R\&D, AstraZeneca}
\icmlaffiliation{comp}{Centre for Misfolding Diseases, Yusuf Hamied Department of Chemistry, University of Cambridge}
\icmlaffiliation{sch}{Biologics Engineering, Oncology R\&D, AstraZeneca}

\icmlcorrespondingauthor{Talip Uçar}{talip.ucar@astrazeneca.com}

\icmlkeywords{Machine Learning, ICML}

\vskip 0.3in
]



\printAffiliationsAndNotice{}  

\begin{abstract}
We investigate the potential of patent data for improving the antibody humanness prediction using a multi-stage, multi-loss training process. Humanness serves as a proxy for the immunogenic response to antibody therapeutics, one of the major causes of attrition in drug discovery and a challenging obstacle for their use in clinical settings. We pose the initial learning stage as a weakly-supervised contrastive-learning problem, where each antibody sequence is associated with possibly multiple identifiers of function and the objective is to learn an encoder that groups them according to their patented properties. We then freeze a part of the contrastive encoder and continue training it on the patent data using the cross-entropy loss to predict the humanness score of a given antibody sequence. We illustrate the utility of the patent data and our approach by performing inference on three different immunogenicity datasets, unseen during training.
Our empirical results demonstrate that the learned model consistently outperforms the alternative baselines and 
establishes new state-of-the-art on five out of six inference tasks, irrespective of the used metric.

\end{abstract}

\section{Introduction}
\label{sec:intro}

Proteins are naturally occurring large molecules, composed of one or more chains of amino acids. They are present in all living organisms and include many essential biological compounds such as enzymes and antibodies. An antibody is a type of protein that binds specifically to proteins called antigens, typically found on the surface of pathogens such as bacteria, viruses, and other foreign substances, including cancer cells. Immune system protects the body from them by having a large number of antibodies in circulation throughout the body, capable of binding specifically to different antigens. Monoclonal antibodies are manufactured in laboratories as clones or copies of a single antibody. They are a significant class of biopharmaceuticals, often used for therapeutic purposes, and have contributed significantly to the overall sales of biopharmaceutical products ~\cite{ecker2015}. Therapeutics based on monoclonal antibodies are typically produced from non-human sources and can lead to an undesired immune response, referred as \emph{immunogenicity}, which is one of the major causes of attrition in the development of monoclonal antibody therapeutics \citep{vandivort2020regulatory}. Thus, many therapeutic monoclonal antibodies often go through a \emph{humanization} process, which involves modifying the antibody to make it more similar to human antibodies, reducing the likelihood of triggering an immune response. This process must be able to identify the humanness of an antibody with little to no errors as there are significant costs associated with attrition in the later stages of the drug development process~\cite{marks2021humanization}.

Motivated by the importance of antibody humanness prediction, we propose a weakly-supervised contrastive learning approach designed to extract informative and contextually rich representations of amino acids and antibodies from patented antibody database (PAD). As the contrastive learning is a well-known instance of self-supervised learning, we refer to our approach as SelfPAD.
The patented antibody database is a rich and diverse source of information with functional characterization of antibodies, consisting of approximately $300,000$ unpaired sequences across seven different species~\cite{krawczyk2021data}. Our approach to leverage the PAD for representation learning and immunogencity prediction departs from conventional approaches, pre-dominantly focused on leveraging information from Observed Antibody Space (OAS). The latter is the database with more than billion different antibody sequences sourced from more than $80$ different studies~\cite{olsen2022observed}. What lacks in the OAS is functional characterization of antibodies which limits representation learning to self-supervised approaches focused on sequence completion tasks \citep{ruffolo2021deciphering, olsen2022ablang, shuai2023iglm}. In contrast, the PAD offers noisy association of sequences with function via more than $16,000$ different patent filings that can be translated into more than $1,000$ tasks, one for each set of potential targets (please see Section~\ref{data_source_preprocessing}). This allows for multi-stage, multi-loss supervised learning, where a model is optimized with two different loss functions during the process. More specifically, we devote the initial stage to representation learning aimed at extracting an effective representations of amino acids. We then switch the loss function to focus the final epochs entirely on a predictive task of interest, e.g., humanness prediction.

We pose the initial stage focused on representation learning from patents as a weakly-supervised contrastive-learning problem, where each antibody sequence is associated with possibly multiple noisy identifiers of function (i.e., potential targets) and the objective is to learn a latent representation that groups them according to their patented objectives. In the first instance, this allows for learning of an antibody representation directly by the contrastive encoder that is informed pre-dominantly by the specific binding to different targets. One of our underlying assumptions is that the amino acid embeddings acquired through our proposed framework carry information about the functional and developability traits of antibodies as the patent filing is typically representative of the drug discovery process where the initial goal is to identify a potent binder, followed by subsequent improvements focused on refining its developability properties, encompassing factors such as immunogenicity, aggregation, viscosity, etc. To demonstrate the utility of patent data for improving the immunogenicity prediction, we continue training the contrastive encoder to predict the humanness score of a given antibody sequence. We explore two different avenues for achieving transfer learning between the stages, the first via the pre-trained encoder itself and the second one via the learned embeddings of individual amino acids. 
During this stage of the training process, we also add a multi-layer perceptron (MLP) on top of the contrastive encoder and continue training on the patent data using the cross-entropy loss function. We evaluate the effectiveness of the learned immunogenicity model by performing inference on three different immunogenicity datasets published as part of prior work (see Sections ~\ref{related_works} and ~\ref{data_source_preprocessing}). Our empirical analysis demonstrates that the final model learned from patent data achieves the new state-of-the-art performance on five out of six inference tasks, irrespective of the used metric.

\section{Related Work}\label{related_works}

\begin{figure*}
     \centering
     \begin{subfigure}[b]{0.55\textwidth}
         \includegraphics[width=1\textwidth,center]{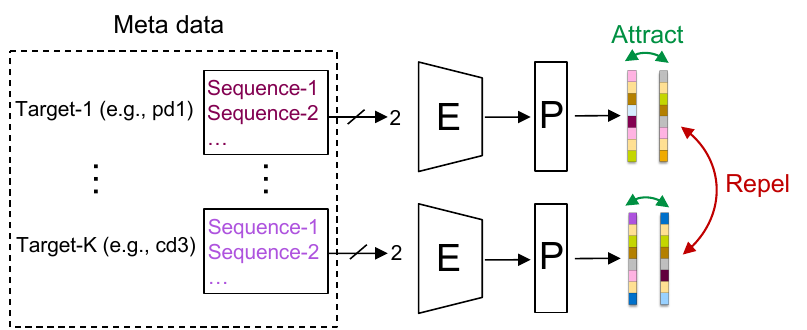}
        \caption{Pre-training}
     \end{subfigure}
     \begin{subfigure}[b]{0.22\textwidth}
         \includegraphics[width=1\textwidth,center]{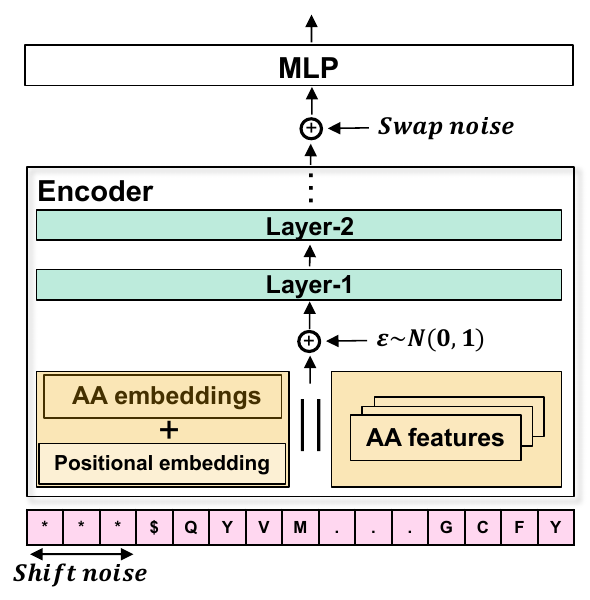}
        \caption{Fine-tuning}
     \end{subfigure}
        \caption{\textbf{SelfPAD Framework:} \textbf{a)} \textbf{Pre-training:} The training process that considers sequences associated with the same potential targets as positive pairs. \textbf{b)} \textbf{Fine-tuning:} The architecture of the transformer-based sequence encoder used in pre-training as well as MLP, swap noise and shift noise used only during fine-tuning.}
        \label{fig:method_selfpad}
\end{figure*}

Various \emph{humanness} scores have been developed to measure the human-like characteristics of humanized antibody sequences and identify potential risks for immunogenicity. Among these, two of the most prominent methods determine humanness by considering sequence identity with a reference library of human sequences, either averaged across all sequences~\citep[Z-score,][]{abhinandan2007analyzing} or across the closest $20$ sequences~\citep[T20 score,][]{gao2013monoclonal}. Both of these approaches provide a single score for the entire sequence and rely on small reference libraries, restricting the diversity of designed therapeutics. Recently, novel methods incorporating machine learning (ML) have been proposed that leverage patterns associated with human and non-human sequences to predict the humanness of an antibody. The most popular approaches along these lines are Hu-mAb~\cite{marks2021humanization}, OASis~\cite{prihoda2022biophi}, and AbNatiV~\cite{Ramon2024}.

\textbf{OASis} is an approach quantifying the humanness of an antibody sequence based on peptide $k$-mers. It works by first segmenting a sequence into overlapping $9$-mer peptides and then querying these fragments against the Observed Antibody Space (OAS)~\cite{olsen2022observed} to estimate their prevalence within the human population. To comprehensively evaluate and compare humanness at the entire antibody level, it introduces the \emph{OASis identity score}, which is calculated at the antibody level as the fraction of its $k$-mer peptides that surpass a user-defined prevalence threshold, where the threshold dictates the required fraction of the human population for a given peptide to be deemed human. OASis offers four predefined thresholds, capturing different stringency levels: \emph{i}) loose ($\geq 1\%$ subjects), \emph{ii}) relaxed ($\geq 10\%$ subjects), \emph{iii}) medium ($\geq 50\%$ subjects), and \emph{iv}) strict ($\geq 90\%$ subjects). For instance, at the relaxed threshold, a peptide is classified as human if found in at least $10\%$ of subjects. Accordingly, the OASis relaxed identity score for an antibody is computed as the fraction of its peptides found in at least $10\%$ of subjects. It is important to note that OASis may not perform effectively for antibodies with long CDR3 loops \cite{prihoda2022biophi}, since these are highly varied regions.

\textbf{AbNatiV} is based on a vector-quantized variational auto-encoder (VQ-VAE)~\cite{van2017neural} trained with masked unsupervised learning on aligned variable domain (Fv) sequences from curated native immune repertoires. Four distinct models are trained on Fv sequences corresponding to \emph{i}-\emph{iii}) human heavy (VH), kappa (VKappa) and lambda (VLamda) chains, and \emph{iv}) camelid heavy chain single-domains (VHH or nanobodies). It is specifically designed for quantifying the humanness of antibodies and nanobodies. Trained on curated datasets of human antibodies or camelid nanobodies, it predicts the likelihood of a given sequence belonging to the distribution of immune-system-derived antibodies. It can be used for both designing antibodies and nanobodies, and predicting the likelihood of immunogenicity. One potential shortcoming of AbNatiV is that it relies on four separate models, one for each chain type, and, thus, fails to account for the entirety of information required for quantifying the humanness property.

\textbf{Hu-mAb} is another approach that utilizes the OAS database and trains multiple random forest (RF) classifiers to distinguish between human and non-human sequences, each tailored for a specific human V gene type. It can be used to infer the \emph{humanness} of sequences as well as for humanization of sequences. Similar to AbNativ, relying on multiple models is one of its potential shortcomings. It is also known to perform worse with light chains~\cite{marks2021humanization}.

When it comes to machine learning aspects of our work, it is related to setups involving self-supervised pre-training and task-specific fine-tuning. The paradigm of pre-training, whether supervised or unsupervised, and subsequent fine-tuning has proven to be a potent approach with promising results across various domains such as natural language processing~\cite{mikolov2013efficient, pennington2014glove, devlin2018bert, liu2019roberta, conneau2019unsupervised, joulin2016bag, collobert2008unified}, computer vision and audio domains~\cite{chen2020simple, grill2020bootstrap, oord2018representation, caron2020unsupervised, he2020momentum, falcon2020framework, chen2020improved}. This progress was mainly enabled by taking advantage of spatial, semantic, or temporal structure in the data through data augmentation~\cite{chen2020simple}, pretext task generation~\cite{devlin2018bert} and using inductive biases through architectural choices (e.g., convolutional layers for images). In this particular instance, we leverage contrastive learning~\cite{chen2020simple} tailored for the multi-task setting at the initial epochs of the training process, and this stage can be considered a pre-training step. The second phase of the training process can be viewed as a form of fine-tuning. Unlike the conventional approach found in the literature, this fine-tuning occurs on the original dataset with the objective of learning to quantify the humanness of an antibody sequence. We should note that a well-curated patent dataset could facilitate the training of a multi-label, multi-task classifier. However, the identifiers of function currently available are noisy and may not be highly reliable. This is often due to a tendency toward risk aversion in safeguarding intellectual property, potentially resulting in patent filings containing sequences with varying functions. An alternative strategy for pre-training might involve disregarding the function and concentrating on self-supervised learning, which typically involves sequence completion tasks. We do not pursue this, however, and leave it as a potential avenue for future exploration. Finally, it's important to emphasize that patent data serves as a unique information source for assessing the humanness of antibodies. Existing works, including those discussed here, predominantly rely on the OAS database. However, it is essential to note that the OAS database exhibits bias and lacks the necessary diversity to comprehensively characterize the entire population of human antibodies.

\section{Method}

In this section, we describe our approach for learning an effective model for predicting the humanness of an antibody sequence. As described in Section~\ref{sec:intro}, we perform a multi-stage training process with two different loss functions. There are two main steps characterizing our approach: \emph{i}) learning an antibody sequence encoder by pre-training on patent data using weakly supervised constrastive learning, and \emph{ii}) learning an antibody property prediction model by fine-tuning the encoder on the same dataset using the cross-entropy loss function and humanness as the task. 

\textbf{Pre-training.} The encoder consists of two components: \emph{i}) a core transformer-based encoder that performs one-to-one mapping between amino acids from the antibody sequence and their context-dependent latent space embeddings and \emph{ii}) a projection layer that maps the latent space representation of an antibody sequence to a fixed dimensional vector, independent of the sequence length. The transformer block has four layers and each one is implemented such that the layer normalization is applied prior to the self-attention and feedforward blocks~\citep[e.g., see][]{radford2019language}. We use a maximum sequence length of $180$ and the dimension of $32$ for the embedding and hidden layers. At the input, we first encode each amino acid in the sequence by using two trainable embedding layers, based on amino-acid type and position in the sequence, i.e., its positional encoding. We sum these two embedding vectors and concatenate the resulting embedding with additional $18$ features that are pre-computed based on the biophysical properties of amino acids such as their hydrophobicity, charge, volume etc. (please see Table~\ref{aa_properties} in the appendix for the full list of properties). We also inject a $\sigma$-scaled Gaussian noise (with $\sigma=0.01$) to the embeddings to learn a robust amino acid representation during training. We do not use any alignment for the sequences and do not explicitly condition the model with respect to the type of chain or the origin of species, i.e., we use each sequence as given. Since we use unigrams, the vocabulary size is twenty three to account for twenty amino acids as well as the tokens used for beginning and end of the sequence, and padding (\#, \$, and * respectively). The projection layer includes a linear layer, followed by GELU activation and another linear layer, where both of them have a dimension of $128$. We pre-trained the model with a batch size of $100$ for $1000$ epochs (see Figure~\ref{fig:pad_loss} in the appendix). 

Figure~\ref{fig:method_selfpad}a depicts our contrastive learning setup. Antibody sequences are grouped by their association with potential function identifiers as specified in patents. Let \( K \) be the total number of targets available. We uniformly sample \( k \) targets without replacement from this set in every iteration during training, i.e.,
\begin{eqnarray}
T=\{t_1,..., t_k\}, \quad t_i \sim \mathcal{U}(1, K), \quad i = 1,.., k \label{eq1},
\end{eqnarray}
where \( t_i \) denotes the index of the \( i \)-th sampled target. This process ensures that each element in the set of \( K \) targets has an equal probability of being selected, resulting in the creation of a representative subset \( T \) of size \( k \). Then, we uniformly sample two sequences (positive pairs) for each of $k$ targets, resulting in a minibatch size of $2k$:
\begin{eqnarray}
S=\{(s_{i1}, s_{i2}) \mid s_{i1}, s_{i2} \sim \mathcal{U}(G_{t_i}), i = 1,.., k\} \label{eq2},
\end{eqnarray}
where $(s_{i1}, s_{i2})$ represents a pair of sequences from a group of sequences $G_{t_i}$ associated with the target $t_i$. This approach ensures that for each target $t_i$, two sequences are uniformly chosen, thereby generating a set of representative samples. Given a positive pair of sequences $(s_{i1}, s_{i2})$, we treat the remaining $2(k - 1)$ sequences within a minibatch as negative examples. As shown in Figure~\ref{fig:method_selfpad}a, we obtain embeddings $(z_{i1}, z_{i2})$ for each pair of sequences by using a shared encoder and a projection network. Then, we compute the normalized temperature-scaled cross entropy loss \cite{chen2020simple} for a positive pair $(\bm{z}_{i1}, \bm{z}_{i2})$ as
\begin{align}
\label{eq4}
\begin{aligned}
& \mathcal{L}_c = \frac{1}{2k}\sum_{i=1}^{k}\left[ l(\bm{z}_{i1}, \bm{z}_{i2}) + l(\bm{z}_{i2}, \bm{z}_{i1})\right] \quad \text{with} & \\
& l(\bm{z}_{i1}, \bm{z}_{i2}) = -\log\frac{\exp(\psi(\bm{z}_{i1},\bm{z}_{i2})/\tau)}{\sum_{j=1, j\neq i}^{2k} \exp(\psi(\bm{z}_{i1}, \bm{z}_{j2})/\tau)}  &
\end{aligned}
\end{align} 
where $\psi(.)$ is the inner product similarity score, $\mathcal{L}_c$ is total contrastive loss, and $l(\bm{z}_{i1}, \bm{z}_{i2})$ is the loss function for a positive pair $(\bm{z}_{i1}, \bm{z}_{i2})$.

\begin{figure*}[th]
     \centering
     \begin{subfigure}[b]{0.28\textwidth}
         \includegraphics[width=1\textwidth,center]{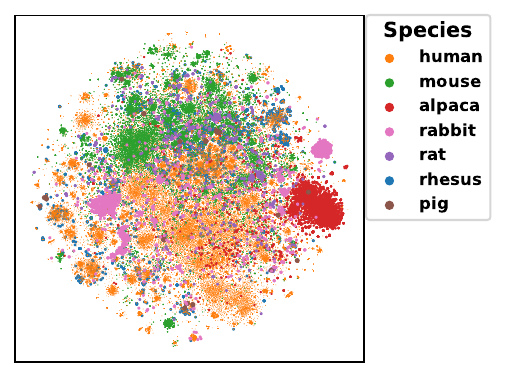}
         \caption{Species}
     \end{subfigure}
     \begin{subfigure}[b]{0.35\textwidth}
         \centering
         \includegraphics[width=1\textwidth,center]{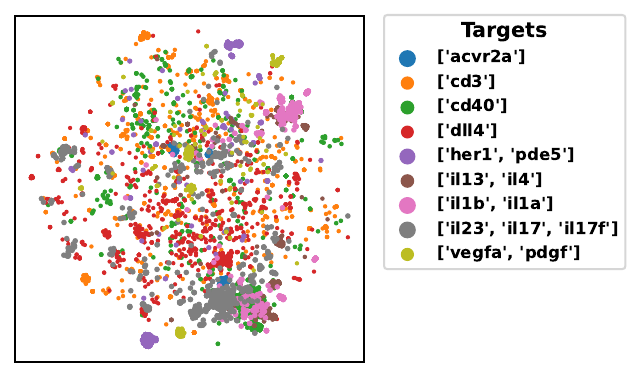}
         \caption{Potential targets}
     \end{subfigure}
     \begin{subfigure}[b]{0.32\textwidth}
         \includegraphics[width=1\textwidth,center]{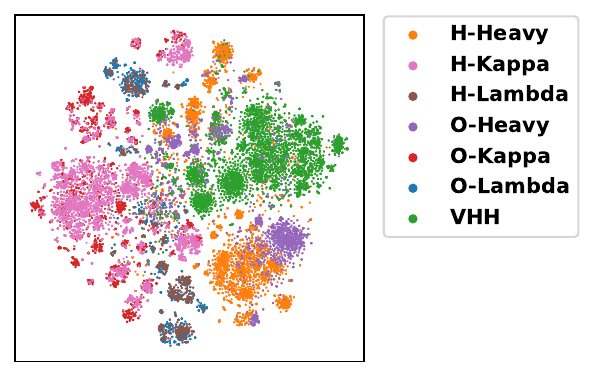}
        \caption{Clusters of chain types}
     \end{subfigure}
        \caption{\textbf{After pre-training:}  t-SNE visualisation of antibodies in the test set based on their \textbf{a)} origin and \textbf{b)} potential targets (illustration for a small subset of targets). \textbf{After fine-tuning the model to predict for humanness:} \textbf{c)} t-SNE plot of heavy, kappa, lambda and VHH chains from different species (H denotes human and O heavy chains from other species). }
        \label{fig:pad_umap}
\end{figure*}

\textbf{Fine-tuning.} For this stage, we add a multi-layer perceptron (MLP) on top of the transformer-based encoder learned in the pre-training step. MLP consists of blocks with a residual connection, where each one has a linear layer with dimension $128 \times 512$ followed by GELU activation, batch normalisation, dropout (with $p=0.3$) and another linear layer with dimension $512 \times 128$. The final linear layer is paired with the softmax activation to give a binary classifer. To learn a robust representation of amino acids and antibodies, we also apply two types of noise during training: \emph{i}) \textit{shift noise}: half of the time, we shift the starting position of a sequence by a randomly sampled number from range $[0, 10]$, and add the mask tokens at the beginning. 
\emph{ii}) \textit{swap noise}: before inputting the resulting embedding into the MLP, we incorporate swap noise \citep{ucar2021subtab}, which involves randomly replacing selected embedding entries with values sampled from the same column.

We explored different strategies for the fine-tuning phase of the training process, where for example, one re-initializes the final four layers of encoder (effectively keeping only the learned amino acid embeddings and position encoding fixed) and fine-tunes by applying different learning rates to each layer of the contrastive encoder as well as the MLP blocks. More specifically, we increase the learning rate from low $0.0001$  to high $0.001$ as we move from lower to upper layers in the encoder. Alternatively, we also investigated the performance of the model that continues training of the contrastive encoder (i.e., no re-initialization of layers) on the humanness task during the fine-tuning epochs (including the extra MLP block). In the second phase of the training, we used cross-entropy loss with label smoothing (configured to $0.5$) and a batch size of $512$ for $25$ epochs. The code for SelfPAD is available at: \url{https://github.com/AstraZeneca/SelfPAD}

\section{Experiments}

In this section, we present our empirical results and provide evidence for the effectiveness of the approach in improving the humanness prediction of antibodies. We start with a high-level description of the used datasets and data-processing performed prior to training. Then, we summarize our empirical observations for the pre-training and fine-tuning steps of the multi-stage training process and illustrate the effectiveness of the approach relative to baselines.

\subsection{Data Sources and Pre-processing}
\label{data_source_preprocessing}

\textbf{Patent data} for antibodies can be a potentially valuable source of information for learning useful representations from their sequences. The primary purpose of patents is to provide legal protection rather than to serve a scientific knowledge, hence such database can be noisy. For example, although patented antibodies are tagged with potential targets, their relationship is not always clear. However, despite of its low signal-to-noise ratio, it can contain valuable information on relationships between family of antibody sequences and their potential targets. To illustrate its utility, we use patented antibody database (PAD) \cite{krawczyk2021data} that contains $16,526$ patent families from major jurisdictions such as US Patent and Trademark Office and World Intellectual Property Organization. These families hold $\sim 290k$ unique antibody chains (unpaired heavy and light chains as well as nanobodies) that are compiled in the PAD. Sequences in the database are manually tagged with potential target(s). If a sequence is associated with more than one target (e.g., \emph{'il13'} and \emph{'il4'}), we use the set of targets as the label (e.g., \{\emph{'il13'}, \emph{'il4'}\}). We group sequences based on their association with potential targets. Sequences in each group constitute our positive samples for the corresponding target. We then randomly split the sequences into training ($\sim 260k$) and test folds/sets ($\sim 29k$) by using stratified sampling to keep the proportion of antibodies from humans and non-humans the same in both sets. Since the PAD also includes sequences from other publicly available datasets, we excluded $553$ therapeutics and $217$ immunogenicity data \cite{prihoda2022biophi}, along with $25$ humanization data \cite{marks2021humanization}, from the training set to avoid data leakage, considering that we utilize these datasets for evaluation purposes. Additionally, we excluded all sequences exhibiting at least $95\%$ sequence similarity with any sequence in the latter three datasets. The minimum Levenshtein distance of sequences in the three evaluation datasets from those in the training set of the PAD is given in Table~\ref{Tab:levenshtein_distance}. During pre-training, we also drop any sequence that is not tagged with any target as well as any target that has less than 20 sequences in its group. After filtering, we end up with $1,063$ potential targets and $\sim 136k$ sequences in the training set, which we then use to pre-train the encoder. During fine-tuning of the pre-trained encoder for humanness prediction, we utilize all the available training set sequences. The training set is then split into two folds, $90\%$ for training and $10\%$ for validation, and the model is fine-tuned with cross-entropy loss for $25$ epochs.

\begin{table}[h]
\centering
\vspace{1mm} 
\caption{Minimum Levenshtein distance between evaluation datasets and the PAD training set. Abbreviations are; \textbf{LD:} Levenshtein Distance, \textbf{553 T-H/T-L:} 553 Therapeutics data Heavy/Light chains, \textbf{217 Im.-H/Im.-L:} 217 Immunogenicity data Heavy/Light chains, \textbf{25 Hum.-P/Hum.-H} 25 Parental/Humanized sequences from humanization data.
\label{Tab:levenshtein_distance}}
\resizebox{0.48\textwidth}{!}{
{\begin{tabular}{@{}c|c|c|c|c|c|c@{}}
\toprule {} & {\bf 553 T-H}& {\bf 553 T-L} & {\bf 217 Im.-H}& {\bf 217 Im.-L} & {\bf 25 Hum.-P}& {\bf 25 Hum.-H} \\\hline \hline
{ \bf LD} & {8} & {6}   & {4}& {8} & {6} & {9}   \\

\hline
\end{tabular}}{}
}\vspace{-1mm}
\end{table}

\begin{figure*}[t]
     \centering
     \begin{subfigure}[b]{0.29\textwidth}
         \centering
         \includegraphics[width=1\textwidth,center]{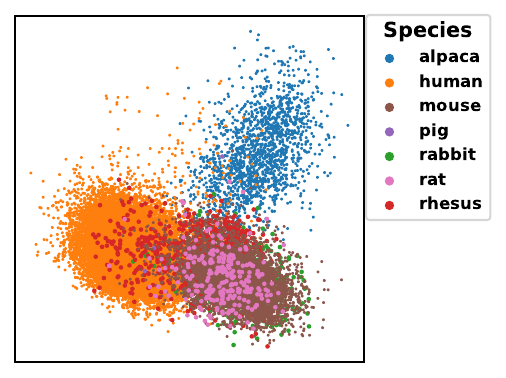}
         \caption{PAD test set}
     \end{subfigure}
     \begin{subfigure}[b]{0.22\textwidth}
         \includegraphics[width=1\textwidth,center]{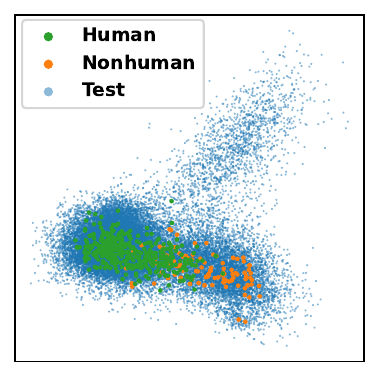}
        \caption{553 therapeutics}
     \end{subfigure}
     \begin{subfigure}[b]{0.22\textwidth}
         \centering
         \includegraphics[width=1\textwidth,center]{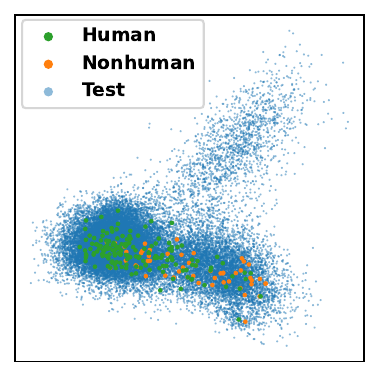}
         \caption{217 ADA data}
     \end{subfigure}
     \begin{subfigure}[b]{0.22\textwidth}
         \includegraphics[width=1\textwidth,center]{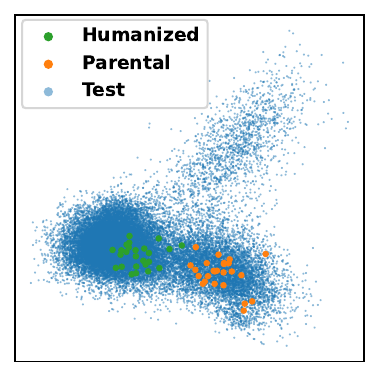}
        \caption{25 humanization data}
     \end{subfigure}
        \caption{\textbf{Human versus Nonhuman sequences:} PCA of embeddings from FT-SelfPAD for \textbf{a)} the test set of PAD, \textbf{b)} 553 therapeutics data from OASis, \textbf{c)} 217 immunogenicity data from OASis, \textbf{d)} 25 humanization data from Hu-mAb, where 25 therapeutic antibodies with the immunogenicity problem (i.e., parental sequence) as well as their humanized versions were plotted together with other test fold sequences from the PAD (shown in blue).} 
        \label{fig:humanness_ada_plots}
\end{figure*}

\textbf{553 Therapeutics} is a dataset from~\citet{prihoda2022biophi}, also used in OASis, and originally sourced from IMGT mAb DB~\cite{poiron2010imgt}. After removing four antibodies (letolizumab, lulizumab pegol, placulumab and glofitamab), we are left with $449$ paired sequences, consisting of $197$ human, $226$ humanized, $63$ chimeric, $41$ humanized/chimeric, $13$ murine, $6$ caninized, and $3$ felinized sequences. For humanness prediction evaluation, we classify the positive class as containing human and humanized sequences ($423$ sequences), while the others are labeled as negative.

\textbf{217 immunogenicity} refers to the dataset obtained from \citet{prihoda2022biophi} and also used in OASis. It contains reported ADA responses from $217$ paired sequences and it was originally curated in \citet{marks2021humanization}. We evaluate our model by scoring $217$ therapeutics for humanness. Sequences with immunogenicity score $\geq10$ are classified as nonhuman while the others are labelled as human.

\textbf{25 humanization data} refers to the dataset with
$7$ pairs of parental-humanized sequences that were first curated by \citet{clavero2018humanization} and expanded to $25$ pairs by \citet{marks2021humanization}. It has also been used in the evaluation of Hu-mAb. We use these $25$ experimentally validated pairs to predict reported ADA responses in the form of humanness prediction. We classify parental sequences with immunogenicity issues as non-human and their corresponding humanized versions as human.

\subsection{Empirical Analysis of the Pre-training Stage} 

Figure~\ref{fig:pad_umap} shows a clustering of sequences from the test set of the PAD, based on their origin and potential targets (Panels~\ref{fig:pad_umap}a-b). We see that the sequences from same species as well as those associated with same potential targets tend to cluster together. Note that when training our models, sequence embeddings are generated by aggregating over the amino acids to prevent the model from memorizing the properties that depend on the sequence length (though it might be useful to learn this dependency in some context).

\begin{table*}[tb]
\caption{\textbf{The PAD test set:} Summary of performance for humanness prediction. Please note that the sequences from alpaca are excluded when computing metrics for AbNatiV and Hu-mAb.}
\label{pad_humanness_performance}
\begin{center}
\begin{small}
\begin{sc}
\resizebox{0.63\textwidth}{!}{
{
\begin{tabular}{l|cccccc}
\hline
\multicolumn{1}{c|}{\textbf{Model}} & \textbf{F1} & \textbf{Recall} & \textbf{Precision}  & \textbf{Accuracy} & \textbf{ROC AUC} & \textbf{PR AUC}\\ \hline
\textbf{FT-SelfPAD}         & \textbf{97.59}       & \textbf{97.36}    & \textbf{97.83}     & \textbf{96.58}  & \textbf{99.12}   &\textbf{98.54}       \\
\textbf{AbNativ}               & 91.59 &  90.68     &   92.52     &  87.54 & 91.52    &  96.49 \\
\textbf{OASis}                 & 88.67 &  86.45     &   91.00     &  83.45  & 88.30   &  95.57 \\
\textbf{Hu-mAb}                & 70.35 &  55.22    &   96.87     &  64.72  & 74.83    &  93.01 \\
\end{tabular}}{}
}
\end{sc}
\end{small}
\end{center}
\vskip -0.1in
\end{table*}

\subsection{Empirical Analysis of the Fine-tuning Stage}
A humanness score should be able to distinguish between human antibodies and antibodies from other species. In particular, it should enable doing so for therapeutic antibodies since those are the primary subjects of humanness analysis. The goal of humanness evaluation is to capture and reduce the immunogenicity risk. If antibodies are not structurally similar to natural human antibodies, there is an increased risk of the immune system recognizing them as foreign and producing antibodies against therapeutic antibodies, i.e., "anti-drug antibodies" (ADAs) or "immunogenicity". Therefore, we fine-tune and evaluate SelfPAD to predict the humanness of sequences. We refer to the fine-tuned model as FT-SelfPAD. We should note that, in the machine learning literature, the term \emph{fine-tuning} typically refers to supervised training of a model, which is pre-trained on a large corpus of data (typically using an unsupervised learning algorithm), on a small dataset. In our context, fine-tuning corresponds to further training of pre-trained model on the same dataset, but using a supervised loss instead of a contrastive loss so that the model can learn to distinguish between human and nonhuman sequences. Once trained, we benchmark FT-SelfPAD against three strong baselines, namely Hu-mAb \cite{marks2021humanization}, OASis \cite{prihoda2022biophi} and AbNatiV \cite{Ramon2024}, by using four datasets: \emph{i}) 29k unpaired sequences from the test set of the PAD, \emph{ii / iii}) paired sequences from 553 therapeutics and 217 immunogenicity data \cite{prihoda2022biophi} and \emph{iv}) paired sequences from 25 humanization data \cite{marks2021humanization}. 

\textbf{29k sequences from the test set of PAD.} Once the model is fine-tuned for humanness prediction, we extract embeddings of the test set, and cluster sequences based on their chain type and origin. As shown in Figure~\ref{fig:pad_umap}c, the sequences from different origins and chain types make up distinct clusters. For examples, heavy chains from humans (orange in Figure~\ref{fig:pad_umap}c) cluster together, but are separate from the heavy chains of other species (purple). Similarly, nanobodies (i.e., VHH sequences) constitute a distinct cluster away from the other sequences. This result indicates that our model should be able to distinguish the origin and chain type of a sequence. Moreover, we note that lambda chains from both humans and other species cluster together, hinting that our model might have more difficulty in distinguishing the origin of lambda chains. We also plot sequences based on their origin by using PCA of embeddings in Figure~\ref{fig:humanness_ada_plots}a, in which we clearly see that FT-SelfPAD can distinguish the sequences from humans, alpaca and other species. Table~\ref{pad_humanness_performance} summarizes the performance of FT-SelfPAD and other baselines for humanness prediction and shows that FT-SelfPAD performs the best across all the metrics. We should note that AbNatiV \cite{Ramon2024} has four models, one for each chain type of heavy, lambda, kappa and VHH. Thus, we used AbNatiV's corresponding models to predict the humanness (with a threshold of 0.8) of heavy, kappa and lambda chains from human and other species in the test set after removing VHH sequences. For OASis, we evaluate and compare humanness by using the OASis identity score and prevalence threshold of \emph{'relaxed'}. We evaluated OASis at three different OASis identity threshold: 0.5, 0.7 and 0.8. We obtained the best result by using identity threshold of 0.7, which we also use in the remaining experiments. Finally, Hu-mAb relies on separate classifiers, each of which is trained for each human V gene type. It also computes humanness score based on each chain type, i.e., heavy or light chain. If it identifies any V gene in the sequence as human, we consider the entire sequence as human. Hence, we scored heavy and light chains from humans and other species after removing VHH sequences. In our experiments, we noticed that Hu-mAb performs poorly on light chains from humans, an observation that is also reported in \cite{marks2021humanization}, resulting in a poor performance overall.

\begin{table*}[tb]
\caption{\textbf{553 therapeutics:} Summary of performance for paired sequences (top) \& single chains (bottom). Please note that we consider both human and humanized sequences as human.}
\label{humanness_performance_553}
\begin{center}
\begin{small}
\begin{sc}
\resizebox{0.63\textwidth}{!}{
{
\begin{tabular}{c|l|ccccccc}
\hline
\multicolumn{2}{c|}{\textbf{Model}} & \textbf{F1} & \textbf{Recall} & \textbf{Precision} & \textbf{Accuracy} & \textbf{ROC AUC} & \textbf{PR AUC} \\ \hline
 \parbox[t]{2mm}{\multirow{4}{*}{\rotatebox[origin=c]{90}{Paired}}} & \textbf{FT-SelfPAD}   & \textbf{93.33}  & \textbf{99.29}  & 88.05   & \textbf{89.07}    & \textbf{94.47} &\textbf{98.16} \\
& \textbf{AbNativ}                &  91.03  &    95.98 &   86.57  & 85.43 &     90.02   &  96.73 \\
& \textbf{OASis}                  &  91.97  &    93.91 &   \textbf{90.11}  & 87.34 &     89.48   &  96.31 \\
& \textbf{Hu-mAb}                 &  91.89  &    95.04 &   88.94  & 87.07 &     77.68   &  93.90 \\ \hline\hline
 \parbox[t]{2mm}{\multirow{4}{*}{\rotatebox[origin=c]{90}{Single}}} & \textbf{FT-SelfPAD}   & \textbf{92.00}  & \textbf{93.74}  & \textbf{90.32}   & \textbf{87.43}    & \textbf{91.84} &\textbf{97.02} \\
& \textbf{AbNativ}                &  87.93  &    87.00 &   88.89  & 81.60 &     86.91   & 95.48 \\
& \textbf{OASis}                  &  88.14  &    87.00 &   89.32  & 81.97 &     85.42   & 94.88 \\
& \textbf{Hu-mAb}                 &  62.05  &    47.64 &   88.96  & 55.10 &     63.90   & 88.47 \\

\end{tabular}}{}
}
\end{sc}
\end{small}
\end{center}
\vskip -0.1in
\end{table*}

\begin{table*}[tb]
\caption{\textbf{217 immunogenicity data:} Summary of performance for paired sequences (top) \& single chains (bottom). Similar to \cite{prihoda2022biophi}, we consider any sequence with immunogenicity $\geq10$ as nonhuman and with $<10$ as human.}
\label{humanness_performance_217}
\begin{center}
\begin{small}
\begin{sc}
\resizebox{0.63\textwidth}{!}{
{
\begin{tabular}{c|l|ccccccc}
\hline
\multicolumn{2}{c|}{\textbf{Model}} & \textbf{F1} & \textbf{Recall} & \textbf{Precision} & \textbf{Accuracy} & \textbf{ROC AUC} & \textbf{PR AUC} \\ \hline
 \parbox[t]{2mm}{\multirow{4}{*}{\rotatebox[origin=c]{90}{Paired}}} & \textbf{FT-SelfPAD}   & 87.50  & 89.63  & 85.47   & 80.65    & \textbf{82.86} &\textbf{93.79} \\
& \textbf{AbNativ}                &  87.91  &    \textbf{90.85} &   85.14  & 81.11 &     79.90   &  92.09\\
& \textbf{OASis}                  &  85.20  &  85.98 &  84.43  &  77.42  &  78.78     &  92.32\\
& \textbf{Hu-mAb}                  &  \textbf{88.96}  & \textbf{90.85}  &  \textbf{87.13}  & \textbf{82.95} &  74.67     &  92.45\\ \hline\hline
 \parbox[t]{2mm}{\multirow{4}{*}{\rotatebox[origin=c]{90}{Single}}} & \textbf{FT-SelfPAD}   & \textbf{83.75}  & \textbf{81.71}  & \textbf{85.90}   & \textbf{76.04}    & \textbf{78.94} & 90.44 \\
& \textbf{AbNativ}                &  83.12  &    79.57 &   77.86  & 75.58 &     77.86   &  \textbf{91.25} \\
& \textbf{OASis}                   & 82.94   & 80.79  &  85.21  & 74.88 &  76.86  & 90.75 \\
& \textbf{Hu-mAb}                  & 59.72   & 45.43  & 87.13   & 53.69 &  62.34   & 86.90 \\
\end{tabular}}{}
}
\end{sc}
\end{small}
\end{center}
\vskip -0.1in
\end{table*}

\textbf{553 therapeutics data.} We process the dataset such that 197 human and 226 humanized sequences are labeled as human while the others are labelled as nonhuman with the assumption that humanized sequences are more likely to be similar to human sequences. We evaluate the models under two scenarios: \emph{i}) \textbf{Paired:} We measure the humanness prediction of a paired sequence. If at least one of two chains (light and heavy) in a paired sequence is predicted as human, we consider the paired sequence as human. \emph{ii}) \textbf{Single chain:} We measure the performance of models for each chain. Table~\ref{humanness_performance_553} summarizes the performance of all models, showing that FT-SelfPAD performs the best for both scenarios.

\textbf{217 immunogenicity data.} The dataset contains reported ADA responses from 217 sequences. Similar to \citet{prihoda2022biophi}, we label sequences  with immunogenicity $\geq10$ as nonhuman while the ones with $<10$ as human. However, we should emphasize that there is a difference between humanness and immunogenicity, and our evaluation of humanness is only a proxy in this context. For example, some fully human antibodies given at high dosage for extended periods of time can end up being more immunogenic than humanised antibodies administered once at low dose. Keeping this in mind, we summarize our results in Table~\ref{humanness_performance_217}, showing that FT-SelfPAD gives the best performance for the most important metrics, namely ROC AUC and Precision-Recall (PR) AUC for paired sequences---it generally performs the best across all metrics, except PR-AUC for the single chains.

\textbf{25 humanization data.} The dataset consists of 25 parental sequences with immunogenicity issues and their corresponding humanized versions. We label parental sequences as nonhuman and humanized ones as human. As shown in Table~\ref{therapeutics_performance_25pair}, FT-SelfPAD classifies all sequences correctly. Moreover, we plot both parental sequences and humanized counterparts by using a PCA on the their embeddings from FT-SelfPAD as shown in Figure~\ref{fig:humanness_ada_plots}d. We also overlay the sequences from  the test set of PAD to give some context. By comparing to the Figure~\ref{fig:humanness_ada_plots}a, we see that parental sequences show up in clusters of other species while the humanized sequences appear among the sequences from human origin. Hence, it appears that our model is able to correctly identify the sequences with immunogenicity issues. We made similar comparisons for both 553 therapeutics and 217 immunogenicity data in Figure~\ref{fig:humanness_ada_plots}b-c, resulting in a similar observation.

\begin{table*}[tb!]
\caption{\textbf{25 humanization data:} Summary of performance for 25 sequences with immunogenicity problem and their corresponding humanized versions \cite{marks2021humanization} for both paired sequences (top) \& single chains (bottom).}
\label{therapeutics_performance_25pair}
\begin{center}
\begin{small}
\begin{sc}
\resizebox{0.63\textwidth}{!}{
{
\begin{tabular}{c|l|ccccccc}
\hline
\multicolumn{2}{c|}{\textbf{Model}} & \textbf{F1} & \textbf{Recall} & \textbf{Precision} & \textbf{Accuracy} & \textbf{ROC AUC} & \textbf{PR AUC} \\ \hline
 \parbox[t]{2mm}{\multirow{4}{*}{\rotatebox[origin=c]{90}{Paired}}} & \textbf{FT-SelfPAD}   & \textbf{100}  & \textbf{100}  & \textbf{100}   & \textbf{100}    & \textbf{100} &\textbf{100} \\
& \textbf{AbNativ}                &  92.00  &    92.00 &   92.00  & 92.00 &     99.52   & 99.53 \\
& \textbf{OASis}                  &  96.00  &    96.00 &   96.00  & 96.00 &     98.00   & 97.52 \\
& \textbf{Hu-mAb}                 &  96.15  &    100 &   92.59  & 96.00 &     96.00     & 96.30\\ \hline\hline
 \parbox[t]{2mm}{\multirow{4}{*}{\rotatebox[origin=c]{90}{Single}}} & \textbf{FT-SelfPAD}   & \textbf{100}  & \textbf{100}  & \textbf{100}   & \textbf{100}    & \textbf{100} &\textbf{100} \\
& \textbf{AbNativ}                &  85.71  &    78.00 &   95.12  & 87.00 &     96.12   &  95.98 \\
& \textbf{OASis}                  &  81.72  &    76.00 &   88.37  & 83.00 &     93.14   &  92.36 \\
& \textbf{Hu-mAb}                 &  64.94  &    50.00 &   92.59  & 73.00 &     73.00   &  83.80 \\
\end{tabular}}{}
}
\end{sc}
\end{small}
\end{center}
\vskip -0.1in
\end{table*}

\begin{figure*}[!t]
     \centering
     \begin{subfigure}[b]{1\textwidth}
         \includegraphics[width=1\textwidth,center]{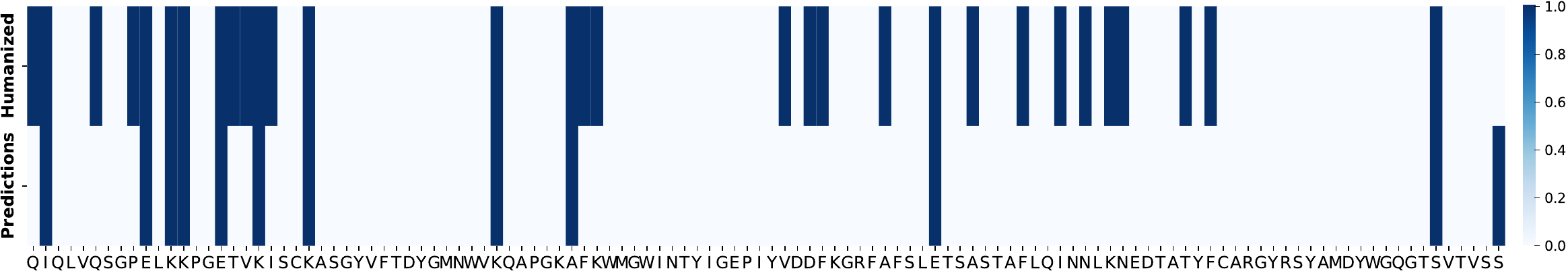}
     \end{subfigure}
        \caption{Comparing SelfPAD-inferred (bottom) to experimental (top) positions in humanizing Certolizumab's heavy chain.} 
        \label{fig:heatmap_certolizumab}
\end{figure*}

\textbf{Interpretability.} 
To enhance the model's utility in antibody engineering, humanness evaluation should offer interpretable scores by highlighting the impact of individual residues or sequence segments. Improved interpretability includes providing explanations of humanness scores through comparisons with closely related reference sequences. We tested FT-SelfPAD's ability to infer positions for humanizing sequences by analyzing attention in 25 humanization data sequences. To identify potential candidates for humanization, we subtracted attention scores (extracted from the last layer of encoder) of humanized sequences from their corresponding parental sequences, scaled the results to $[0,1]$, and applied a threshold of $0.7$ for binarization.  Figure~\ref{fig:heatmap_certolizumab} displays positions with the highest attention, comparing inferred positions with mutated positions in the humanized sequence of Certolizumab (see Section~\ref{interpretability_appendix} in the appendix for more results). The model focuses on positions initially mutated in the parental sequence, recommending fewer mutations than those found in the humanized sequences. This aligns with observations in \cite{marks2021humanization}, suggesting the potential for fewer mutations in humanization.

\textbf{Alignment with biological principles.}
To understand the model's alignment with biological principles, we compared sequences from the PAD's test set and the 217 Immunogenicity dataset, and analyzed global relationships and differences among closely related sequences. Please use Figure~\ref{fig:humanness_ada_plots}a and ~\ref{fig:humanness_ada_plots}c as reference for the following analysis. 

First, we extracted PCA embeddings for human sequences in the PAD test set (orange cluster in Figure~\ref{fig:humanness_ada_plots}a), identifying 200 central sequences—90 light chains and 110 heavy chains. These central sequences were used as reference points. Next, we extracted PCA embeddings for sequences in the 217 Immunogenicity dataset (Figure~\ref{fig:humanness_ada_plots}c). We then aligned the heavy chains and separately aligned the light chains from both datasets using the Kabat numbering scheme. This allowed us to compare sequence similarities across datasets. We calculated maximum and average sequence similarities between correctly classified, misclassified, false positive, and false negative sequences in the 217 dataset and the central sequences from the PAD test set. Table~\ref{table:model_alignment} summarizes this analysis with the maximum and average similarities between sequences in the 217 dataset and the central sequences from the PAD test set. As we move from left to right in the table, the sequence similarity decreases, illustrating global sequence relationships.

Our findings show that correctly predicted sequences have higher maximum and average similarity to the central sequences compared to the misclassified ones. False positives also exhibit higher similarities than false negatives. This demonstrates a meaningful correlation between a sequence's Euclidean distance from the center of the human cluster in PCA space and its sequence similarity to the central sequences. However, it is important to note that while there is a correlation between sequence similarity and PCA-derived Euclidean distance, the model's predictions are based on specific sequence motifs, not just overall sequence similarity as shown in Figure~\ref{fig:heatmap_certolizumab} and~\ref{fig:heatmap3_0.8_herceptin_bevacizumab}.

\textbf{Ablation study.}
We performed an ablation study on the 217 immunogenicity dataset to analyze design choices in FT-SelfPAD. Table~\ref{table:ablation_217} shows that incorporating shift and swap noise, along with re-initializing four pre-trained model layers before fine-tuning, improves performance. Notably, the model is less effective when trained from scratch or fine-tuned without random initialization of its four layers (see the first and second rows in Table~\ref{table:ablation_217}).

\begin{table}[h]
\centering
\caption{Comparing maximum and average similarity between the sequences in 217 dataset and the sequences at the centre of human sequences in the test set of PAD. \textbf{TP/FP:} True/False positive, \textbf{TN/FN:} True/False negative.}
\resizebox{0.4\textwidth}{!}{
{\begin{tabular}{@{}l|c|c|c|c|c@{}}
\toprule 
{\bf Similarity} & {\bf Chain Type} & {\bf TP} & {\bf FP} & {\bf FN}   & \textbf{TN} \\\hline \hline
 \parbox[t]{2mm}{\multirow{2}{*}{\rotatebox[origin=c]{0}{Maximum}}} & {Light}& {91.74}  & {89.25} & {79.33}  & {75.21} \\
& {Heavy}& {82.39} & {83.80} & {80.28}  & {70.42} \\
\hline
 \parbox[t]{2mm}{\multirow{2}{*}{\rotatebox[origin=c]{0}{Average}}} & {Light}& {56.50}  & {61.75} & {55.56}  & {53.59} \\
& {Heavy}& {54.49} & {59.26} & {55.46}  & {47.30} \\
\hline
\end{tabular}}{}
}\vspace{-0.11in}
\label{table:model_alignment}
\end{table}

\begin{table}[h]
\centering
\caption{Ablation study on 217 immunogenicity data. \textbf{Swap}: Swap noise, \textbf{Shift}: Shift noise, \textbf{RI}: Four pre-trained model layers randomly initialized before fine-tuning, \textbf{FS}: Training from scratch without pre-training.}
\resizebox{0.4\textwidth}{!}{
{\begin{tabular}{@{}c|c|c|c|c|c@{}}
\toprule {\bf FS} & {\bf RI} & {\bf Swap} & {\bf Shift}   & \textbf{ROC AUC}  & \textbf{PR AUC}\\\hline \hline
{ + }& {NA}  & { + } & { + }  & 80.32 & 92.32\\
{NA}&  { - } & { + } & { + }  & 81.90 & 93.33\\
{NA} & { - } & { - } & { - }  &  80.74 & 91.89 \\
{NA} & { + } & { - } & { - }  &  81.53 & 92.21\\
{NA} & { + } & { + } & { - }  & 82.37  & 93.54\\
{NA} & { + } & { + } & { + }  & \textbf{82.86} & \textbf{93.79}\\
\hline
\end{tabular}}{}
}\vspace{-0.1in}
\label{table:ablation_217}
\end{table}

\section{Conclusion}
In this work, we present a framework, SelfPAD, to learn representations from the Patented Antibody Database, and demonstrate its usefulness by providing a robust model for humanness prediction - a practical challenge in therapeutic antibody development. SelfPAD opens new avenues for advancing the understanding and optimization of antibody sequences, ultimately facilitating the development of more effective and safe therapeutic interventions. Despite its advantages, one of the main drawbacks of our approach is that the PAD is a noisy dataset and further curation is needed to improve the signal-to-noise ratio. Moreover, the PAD introduces new biases, e.g., it biases towards lab optimised rather than natural sequences. Thus, we can also improve our results by incorporating data from other sources such as the OAS since it includes a large number of natural sequences. Finally, although this work focuses on humanness prediction, we can fine-tune SelfPAD for other downstream tasks such as predicting developability properties of antibodies.

\section*{Impact Statement}

\textbf{Data Privacy and Intellectual Property Rights.} Our study involves the utilisation of patent data, which raises important considerations regarding data privacy and intellectual property rights. All the data used in our research have been obtained in compliance with the relevant regulations. We also point out that although sequences in the patent data are released and available in the public domain, their use is subject to intellecutal property rights and, thus, they cannot be used directly as a treatment for a particular disease/indication. Finally, the legal landscape regarding copyright and scientific data may evolve, so it is important to stay informed about the developments in this area.

\textbf{Potential Biases in Patent Databases.} By developing predictive models for antibody humanness, we aim to enhance the efficiency and safety of therapeutic antibody development. However, we acknowledge the broader ethical implications of our work and the possibility of biases inherent in patent databases, including selection bias and geographic bias, which may influence the representativeness of the data. The patent dataset includes both natural and engineered antibody sequences. However, the quantity of natural sequences is limited, which might impact the overall performance of the model. One approach to improve the robustness and generalisability of our findings would be to enrich the dataset by including more natural sequences from the Observed Antibody Space (OAS dataset) and other sources.

\textbf{Limitations and Uncertainties.} There are several constraints inherent in using patent data for predictive modelling. These include the inherent biases in the patent filings, the lack of standardised annotations for antibody sequences, and potential inaccuracies in patent databases. Additionally, identifiers of function (potential targets) in the patent dataset are noisy and may not be completely reliable. This is often due to a tendency toward risk aversion in safeguarding intellectual property, potentially resulting in patent filings containing sequences with varying functions. This can potentially corrupt the representation learned during our pre-training stage. We try to make the model robust to noise by employing various techniques as outlined in our response to previous question(s). Finally, we encourage future research to address these limitations through better data collection methods, validation studies, and robustness assessments.

\section*{Acknowledgements}
We thank the anonymous reviewers for their constructive feedback, and appreciate the support from everyone in the MLAB program at AstraZeneca. Special thanks to Massimo Sammito, Owen Vickery, Sridhar Neelamraju, and Isabelle Sermadiras for valuable discussions.

\nocite{langley00}

\onecolumn
\twocolumn

\bibliography{main}
\bibliographystyle{icml2024}

\newpage
\appendix
\onecolumn

\section{Amino Acid Features}\label{appendix_aa_features}

\begin{table*}[!ht]
\caption{18 biophysical features of amino acids used in SelfPAD.}
\label{aa_properties}
\begin{center}
\begin{small}
\begin{sc}
\resizebox{0.3\textwidth}{!}{
{
\begin{adjustbox}{angle=90}
\begin{tabular}{|c|c|c|c|c|c|c|c|c|c|c|c|c|c|c|c|c|c|c|}
\hline
   & Hydropathy                       &                              &                                 & Chemical Properties            &                             &                               &                            &                            &                             &                               & Volume & Charge                        &                                &                               & Hydrogen                   &                               & Polarity                   &                               \\ \hline
   &                                  &                              &                                 &                                &                             &                               &                            &                            &                             &                               &        &                               &                                &                               &                            &                               &                            &                               \\ \hline
   & \multicolumn{1}{l|}{Hydrophobic} & \multicolumn{1}{l|}{Neutral} & \multicolumn{1}{l|}{Hyrophilic} & \multicolumn{1}{l|}{Aliphatic} & \multicolumn{1}{l|}{Sulfur} & \multicolumn{1}{l|}{Hydroxyl} & \multicolumn{1}{l|}{Amide} & \multicolumn{1}{l|}{Basic} & \multicolumn{1}{l|}{Acidic} & \multicolumn{1}{l|}{Aromatic} &        & \multicolumn{1}{l|}{Positive} & \multicolumn{1}{l|}{Uncharged} & \multicolumn{1}{l|}{Negative} & \multicolumn{1}{l|}{Donor} & \multicolumn{1}{l|}{Acceptor} & \multicolumn{1}{l|}{Polar} & \multicolumn{1}{l|}{Nonpolar} \\ \hline
I  & 4.5                              & 0                            & 0                               & 1                              & 0                           & 0                             & 0                          & 0                          & 0                           & 0                             & 166.7  & 0                             & 1                              & 0                             & 0                          & 0                             & 0                          & 1                             \\ \hline
V  & 4.2                              & 0                            & 0                               & 1                              & 0                           & 0                             & 0                          & 0                          & 0                           & 1                             & 140    & 0                             & 1                              & 0                             & 0                          & 0                             & 0                          & 1                             \\ \hline
L  & 3.8                              & 0                            & 0                               & 1                              & 0                           & 0                             & 0                          & 0                          & 0                           & 0                             & 166.7  & 0                             & 1                              & 0                             & 0                          & 0                             & 0                          & 1                             \\ \hline
F  & 2.8                              & 0                            & 0                               & 0                              & 0                           & 0                             & 0                          & 0                          & 0                           & 1                             & 189.9  & 0                             & 1                              & 0                             & 0                          & 0                             & 0                          & 1                             \\ \hline
C  & 2.5                              & 0                            & 0                               & 0                              & 1                           & 0                             & 0                          & 0                          & 0                           & 0                             & 108.5  & 0                             & 1                              & 0                             & 0                          & 0                             & 0                          & 1                             \\ \hline
M  & 1.9                              & 0                            & 0                               & 0                              & 1                           & 0                             & 0                          & 0                          & 0                           & 0                             & 162.9  & 0                             & 1                              & 0                             & 0                          & 0                             & 0                          & 1                             \\ \hline
A  & 1.8                              & 0                            & 0                               & 1                              & 0                           & 0                             & 0                          & 0                          & 0                           & 0                             & 88.6   & 0                             & 1                              & 0                             & 0                          & 0                             & 0                          & 1                             \\ \hline
W* & -0.9                             & 0                            & 0                               & 0                              & 0                           & 0                             & 0                          & 0                          & 0                           & 1                             & 227.8  & 0                             & 1                              & 0                             & 1                          & 0                             & 0                          & 1                             \\ \hline
G  & 0                                & -0.4                         & 0                               & 0                              & 0                           & 0                             & 0                          & 0                          & 0                           & 0                             & 60.1   & 0                             & 1                              & 0                             & 0                          & 0                             & 0                          & 1                             \\ \hline
T  & 0                                & -0.7                         & 0                               & 0                              & 0                           & 1                             & 0                          & 0                          & 0                           & 0                             & 116.1  & 0                             & 1                              & 0                             & 1                          & 1                             & 1                          & 0                             \\ \hline
S  & 0                                & -0.8                         & 0                               & 0                              & 0                           & 1                             & 0                          & 0                          & 0                           & 0                             & 89     & 0                             & 1                              & 0                             & 1                          & 1                             & 1                          & 0                             \\ \hline
Y  & 0                                & -1.3                         & 0                               & 0                              & 0                           & 0                             & 0                          & 0                          & 0                           & 0                             & 193.6  & 0                             & 1                              & 0                             & 1                          & 1                             & 1                          & 0                             \\ \hline
P  & 0                                & -1.6                         & 0                               & 0                              & 0                           & 0                             & 0                          & 0                          & 0                           & 0                             & 112.7  & 0                             & 1                              & 0                             & 0                          & 0                             & 0                          & 1                             \\ \hline
H  & 0                                & -3.2                         & 0                               & 0                              & 0                           & 0                             & 0                          & 1                          & 0                           & 0                             & 153.2  & 1                             & 0                              & 0                             & 1                          & 1                             & 1                          & 0                             \\ \hline
N  & 0                                & 0                            & -3.5                            & 0                              & 0                           & 0                             & 1                          & 0                          & 0                           & 0                             & 114.1  & 0                             & 1                              & 0                             & 1                          & 1                             & 1                          & 0                             \\ \hline
D  & 0                                & 0                            & -3.5                            & 0                              & 0                           & 0                             & 0                          & 0                          & 1                           & 0                             & 111.1  & 0                             & 0                              & 1                             & 0                          & 1                             & 1                          & 0                             \\ \hline
Q  & 0                                & 0                            & -3.5                            & 0                              & 0                           & 0                             & 1                          & 0                          & 0                           & 0                             & 143.8  & 0                             & 1                              & 0                             & 1                          & 1                             & 1                          & 0                             \\ \hline
E  & 0                                & 0                            & -3.5                            & 0                              & 0                           & 0                             & 0                          & 0                          & 1                           & 0                             & 138.4  & 0                             & 0                              & 1                             & 0                          & 1                             & 1                          & 0                             \\ \hline
K  & 0                                & 0                            & -3.9                            & 0                              & 0                           & 0                             & 0                          & 1                          & 0                           & 0                             & 168.6  & 1                             & 0                              & 0                             & 1                          & 0                             & 1                          & 0                             \\ \hline
R  & 0                                & 0                            & -4.5                            & 0                              & 0                           & 0                             & 0                          & 1                          & 0                           & 0                             & 173.4  & 1                             & 0                              & 0                             & 1                          & 0                             & 1                          & 0                             \\ \hline
\end{tabular}\end{adjustbox}
}{}
}
\end{sc}
\end{small}
\end{center}
\vskip -0.1in
\end{table*}

\begin{figure}
     \centering
     \begin{subfigure}[b]{0.5\textwidth}
         \includegraphics[width=1\textwidth,center]{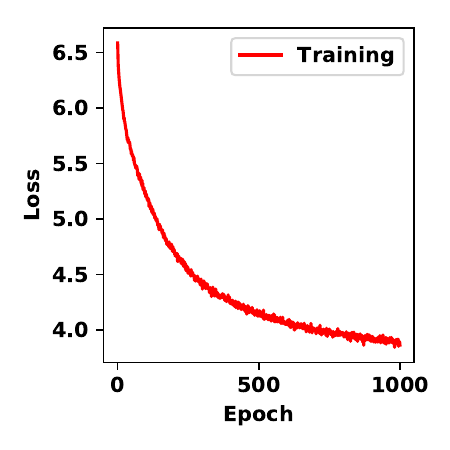}
     \end{subfigure}
        \caption{Training loss during pre-training of SelfPAD.}
        \label{fig:pad_loss}
\end{figure}



\section{Implementation and resources}\label{implementation_resources}
We implemented our work using PyTorch \citep{NEURIPS2019_9015}. AdamW optimizer \citep{loshchilov2017decoupled} with $betas=(0.9, 0.999)$ and $eps=1e-07$ is used for all of our experiments. We used a compute cluster consisting of A10G GPUs throughout this work. For pre-training, we did a hyper-parameter sweep over the parameters such as number of layers of the encoder, number of attention heads, embedding dimension, temperature constant for contrastive loss, learning rate, total epochs etc. and chose the parameters based on the best convergence on the training set. For fine-tuning phase, we did a hyper-parameter search using a validation set.

Table~\ref{hyper_params} lists hyperparameters used for pre-training and fine-tuning of the model during this work while Table~\ref{arch_params} shows the architecture used for the encoder. MLP consists of one block with a residual connection, where the block has a linear layer (128x512) followed by GELU activation, batch normalisation, dropout (p=0.3) and another linear layer (512x128). A final linear layer (128x2) with the softmax activation is used to make a binary classification.

\begin{table*}[!tb]
\caption{\textbf{Hyperparameters:} Hyperparameters used for both pre-training and fine-tuning stages. Abbreviations are; \textbf{LR:} Learning Rate, \textbf{P:} Dropout rate, \textbf{LS:} Label Smoothing used for cross-entropy loss, specifying the amount of smoothing when computing the loss, \textbf{BS:} Batch Size, \textbf{temp.:} Temperature parameter used for contrastive loss, $\bm P_{swap}:$ Percentage of embedding dimension used for swap-noise, $\bm R_{shift}:$ Range used to sample a number that is used to shift the position of sequence to add shift-noise.}
\label{hyper_params}
\begin{center}
\begin{small}
\begin{sc}
\resizebox{0.63\textwidth}{!}{
{
\begin{tabular}{l|c|c|c|c|c|c|c|c}
\hline
                            & \textbf{epoch} & \textbf{bs} & \textbf{lr} & \textbf{p} & \textbf{ls} & \textbf{temp.} & $\bm P_{swap}$ & $\bm N_{shift}$ \\ \hline
\textbf{Pre-training}       & 1000          & 100         & 1e-3          & 0.2   & NA     &  0.1 & NA & NA\\ \hline
\textbf{Fine-tuning}        &  25          & 512      &     [1e-4, 1e-3] &  0.3   & 0.5    & NA  & 0.2 & $[0, 10]$  \\ \hline
\end{tabular}}{}
}
\end{sc}
\end{small}
\end{center}
\vskip -0.1in
\end{table*}

\begin{table*}[!tb]
\caption{\textbf{Encoder architecture:} Abbreviations are; $\bm N_{layers}:$ Number of layers, $\bm N_{heads}:$ Number of attention heads in each layer, $\bm d_{emb}:$ Dimension of embedding in each layer, $\bm d_{head}:$ Feature dimension of each attention head.}
\label{arch_params}
\begin{center}
\begin{small}
\begin{sc}
\resizebox{0.4\textwidth}{!}{
{
\begin{tabular}{l|c|c|c|c}
\hline
                            & $\bm N_{layers}$ & $\bm N_{heads}$ & $\bm d_{emb}$ & $\bm d_{head}$  \\ \hline
\textbf{Encoder}            & 4                & 4                 & 32          & 16     \\ \hline
\end{tabular}}{}
}
\end{sc}
\end{small}
\end{center}
\vskip -0.1in
\end{table*}

\clearpage

\section{Interpretability}\label{interpretability_appendix}

\begin{figure*}[h]
     \centering
     \begin{subfigure}[b]{1.0\textwidth}
         \includegraphics[width=1\textwidth,center]{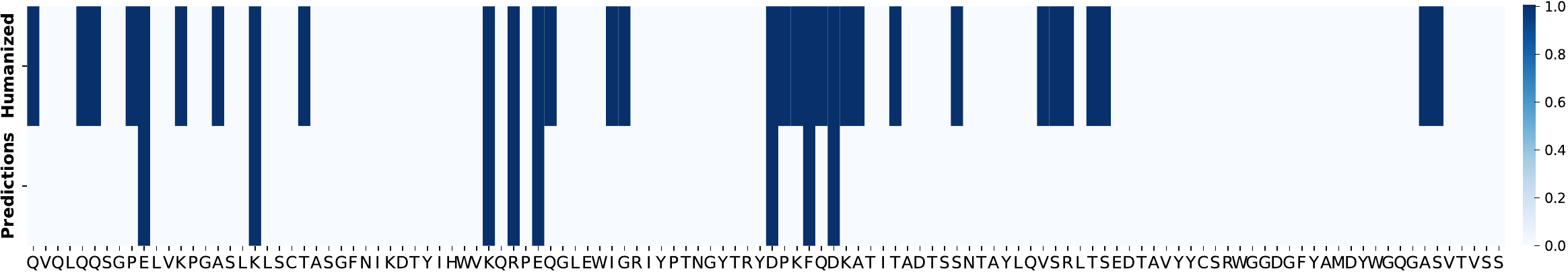}
     \end{subfigure}
     \begin{subfigure}[b]{1.0\textwidth}
         \includegraphics[width=1\textwidth,center]{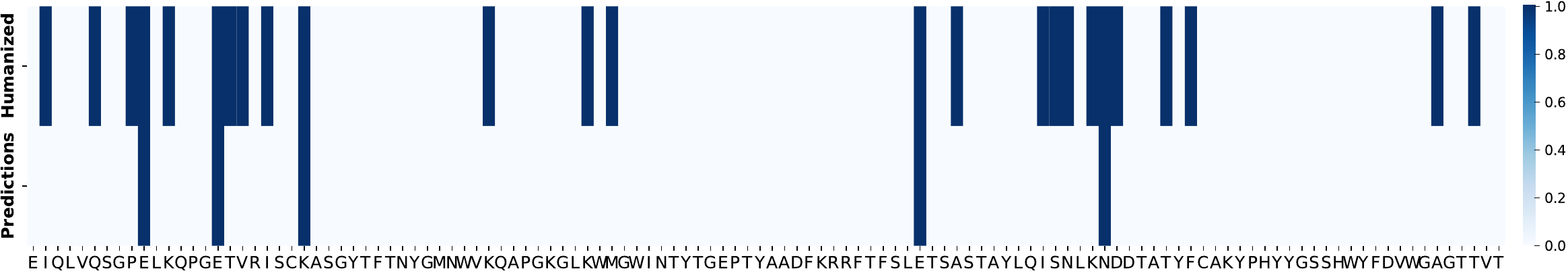}
     \end{subfigure}
     \begin{subfigure}[b]{1.0\textwidth}
         \includegraphics[width=1\textwidth,center]{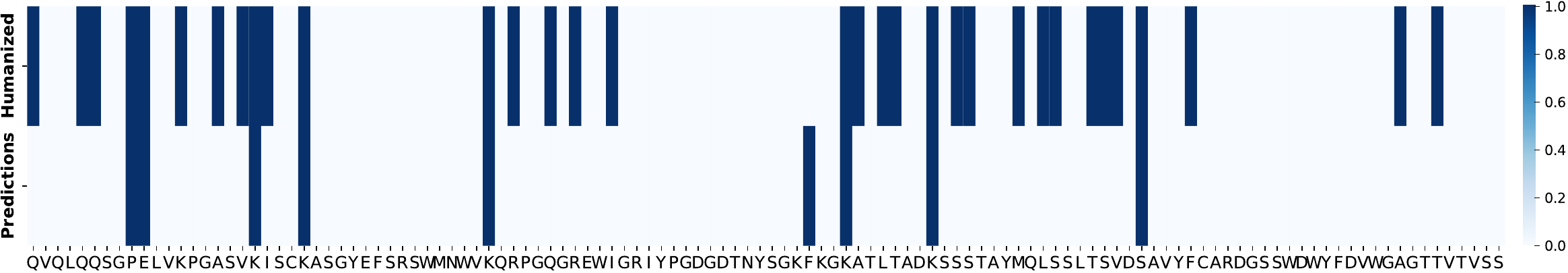}
     \end{subfigure}
        \caption{Comparing SelfPAD-inferred positions to experimental data where the changes are made to humanize the heavy chain of the parental sequences of Herceptin (top), Bevacizumab (middle) and Pinatuzumab (bottom).} 
        \label{fig:heatmap3_0.8_herceptin_bevacizumab}
\end{figure*}

\end{document}